\documentclass[sigconf]{acmart}
\AtBeginDocument{%
  }

\copyrightyear{2025}
\acmYear{2025}
\setcopyright{cc}
\setcctype{by-sa}
\acmConference[CIKM '25]{Proceedings of the 34th ACM International Conference on Information and Knowledge Management}{November 10--14, 2025}{Seoul, Republic of Korea}
\acmBooktitle{Proceedings of the 34th ACM International Conference on Information and Knowledge Management (CIKM '25), November 10--14, 2025, Seoul, Republic of Korea}\acmDOI{10.1145/3746252.3761493}
\acmISBN{979-8-4007-2040-6/2025/11}
\acmISBN{978-1-4503-XXXX-X/2018/06}

\usepackage{xcolor}

\definecolor{topiccolor}{HTML}{1F78B4} %
\definecolor{unicolor}{HTML}{33A02C} %
\definecolor{biunicolor}{HTML}{E31A1C} %
\definecolor{aspectcolor}{HTML}{FF7F00} %
\definecolor{questioncolor}{HTML}{008080} %
\definecolor{filtercolor}{HTML}{A65628} %

\begin{document}

\title{Compare: A Framework for Scientific Comparisons}%

\author{Moritz Staudinger}
\orcid{0000-0002-5164-2690}
\affiliation{%
  \institution{TU Wien}
  \city{Vienna}
  \country{Austria}
}
\email{moritz.staudinger@tuwien.ac.at}

\author{Wojciech Kusa}
\orcid{0000-0003-4420-4147}
\affiliation{%
  \institution{NASK National Research Institute}
  \city{Warsaw}
  \country{Poland}}
\email{wojciech.kusa@nask.pl}

\author{Matteo Cancellieri}
\orcid{0000-0002-9558-9772}
\affiliation{%
  \institution{Open University}
  \city{Milton Keynes}
  \country{United Kingdom}}
\email{matteo.cancellieri@open.ac.uk}

\author{David Pride}
\orcid{0000-0002-7162-7252}
\affiliation{%
  \institution{Open University}
  \city{Milton Keynes}
  \country{United Kingdom}}
\email{david.pride@open.ac.uk}

\author{Petr Knoth}
\orcid{0000-0003-1161-7359}
\affiliation{%
  \institution{Open University}
  \city{Milton Keynes}
  \country{United Kingdom}}
\email{petr.knoth@open.ac.uk}
\author{Allan Hanbury}
\orcid{0000-0002-7149-5843}
\affiliation{%
  \institution{TU Wien}
  \city{Vienna}
  \country{Austria}}
\email{allan.hanbury@tuwien.ac.at}

\renewcommand{\shortauthors}{Staudinger et al.}

\begin{abstract}
Navigating the vast and rapidly increasing sea of academic publications to identify institutional synergies, benchmark research contributions and pinpoint key research contributions has become an increasingly daunting task, especially with the current exponential increase in new publications.
Existing tools provide useful overviews or single-document insights, but none supports structured, qualitative comparisons across institutions or publications.

To address this, we demonstrate \textit{Compare}, a novel framework that tackles this challenge by enabling sophisticated long-context comparisons of scientific contributions.
\textit{Compare} empowers users to explore and analyze research overlaps and differences at both the institutional and publication granularity, all driven by user-defined questions and automatic retrieval over online resources. 
For this we leverage on Retrieval-Augmented Generation over evolving data sources to foster long context knowledge synthesis. Unlike traditional scientometric tools, \textit{Compare} goes beyond quantitative indicators by providing qualitative, citation-supported comparisons.
\end{abstract}




\maketitle

\section{Introduction}
In today's research environment, the unprecedented growth in the number of scientific publications has created a double-edged sword. While access to knowledge has never been as broad and straightforward as it is today, the ability to find, grasp, and understand research outcomes is becoming increasingly strained.
Each year, tens of thousands of new articles are published in journals and conferences~\cite{bornmann_growth_2021}, with entire disciplines undergoing rapid transformation in the production and dissemination of knowledge.
This poses significant challenges for individual researchers and research analysts -- not only requiring them to stay up to date within their own fields of expertise, but also to understand broader developments across institutions and disciplines.

Despite a recent surge in literature discovery and summarization tools, most existing systems are geared toward analyzing individual documents or providing overviews of research topics such as CORE-GPT~\cite{pride_core-gpt_2023}, OpenScholar~\cite{asai_openscholar_2024}, Consensus\footnote{\url{https://consensus.app/}} or Elicit\footnote{\url{https://elicit.com/}}, by applying Retrieval-Augmented Generation~(RAG)~\cite{lewis_retrieval-augmented_2020} to a scientific corpus. 
What remains largely unexplored are tools that support comparative understanding and knowledge synthesis, with the tools only starting to become available in 2025, such as Ai2 by Singh et al.~\cite{singh_ai2_2025} and LitLLM~\cite{agarwal_litllms_2025}. Addressing this gap requires tools that can analyze how research topics, methods, and contributions evolve over time.
This issue becomes especially problematic when researchers aim to identify collaborators, evaluate competing approaches, or understand how their institution’s research profile compares to others.
In the absence of effective comparison and synthesis frameworks, researchers are left to conduct labor-intensive analyses and scoping reviews, or to rely on simplified proxies -- such as institutional rankings or impact factors -- that fail to capture deeper qualitative differences.

To address this challenge, we introduce \textit{Compare}, a modular framework for long-context scientific comparisons at both the institutional and publication levels. To the best of our knowledge, \textit{Compare} is the first tool to combine RAG with long-context synthesis for flexible, question-driven comparisons in the scientific domain. We leverage multiple open-access resources using a two-step RAG pipeline, enabling users to pose custom comparative questions and receive synthesized insights drawn from a large pool of scholarly content. By providing comparisons and statistics based on the user input, \textit{Compare} empowers researchers, research analysts and repository managers to efficiently navigate the growing sea of scientific publications.

\section{Related Work}\label{sec:relatedwork}
\subsection{Scientific Question Answering}
Multi-Document Question Answering~(MDQA) is a well-established research area in both open-domain~\cite{tang_multihop-rag_2024, yang_hotpotqa_2018} and scientific settings~\cite{auer_sciqa_2023, wang_leave_2024}. However, most existing resources focus on rather simple, short-context tasks. To our knowledge, only \citet{wang_leave_2024} explicitly address scientific question answering with long contexts of up to 200,000 characters. Other work, such as \citet{asai_openscholar_2024} and \citet{auer_sciqa_2023}, focuses on multi-document QA but within shorter-contexts.

Query-Focused Multi-Document Summarization~(QMDS) is a related subfield of question answering that aims to summarize multiple documents in response to a user’s information need~\cite{roy_review_2024}. While QMDS overlaps conceptually with our work, existing research in this area remains limited and typically focuses on short documents, such as debates~\cite{nema_diversity_2017} or webpages~\cite{kulkarni_aquamuse_2020, liu_querysum_2024}.

\textit{Compare} builds on these directions but introduces a novel focus on long-context, qualitative comparisons between scientific entities (e.g., institutions or publications), grounded in scholarly sources. Rather than answering factual questions or generating summaries, \textit{Compare} synthesizes similarities and differences using retrieval-augmented generation (RAG) pipelines customized for comparative research analysis.

\subsection{Literature Review Automation}
In recent years, various tools and datasets have been developed to automate scientific literature reviews. Lu et al.~\cite{lu_multi-xscience_2020} introduced the Multi-XScience dataset for abstractive summarization of scientific articles, and Pilault~\cite{pilault_extractive_2020} explored language model-based summarization for scientific texts. Kusa et al.~\cite{kusa_cruise-screening_2023} released \textit{CRUISE-Screening}, a system for automated literature screening from online sources.

More recently, Agarwal et al.~\cite{agarwal_litllms_2025} proposed LitLLM, a structured planning-and-generation pipeline that can generate related work sections from abstracts. Similarly, Singh et al.~\cite{singh_ai2_2025} developed Ai2, a system that supports scientific question answering and knowledge synthesis over the Semantic Scholar corpus.

In contrast to these approaches, our goal is not to generate comprehensive related work sections or article-level summaries. Instead, we provide a flexible pipeline capable of synthesizing information at both the publication and institutional levels, with a focus on comparative insight rather than narrative review construction.
\subsection{Institutional Comparison}
The comparison and evaluation of scientific research has a long history, dating back at least to the establishment of peer review in 1665~\cite{kachooei_editorial_2022}. More systematic approaches such as citation tracking and co-authorship analysis only emerged centuries later. Today, scientometric methods are widely used to assess the impact of research and researchers. For example,~\citet{luo_systematic_2022} performed a scientometric analysis of construction research using co-authorship networks, citation data, and topic clustering. Similarly,~\citet{oyewola_exploring_2022} explored the landscape of machine learning research through scientometric techniques.

Despite their prevalence, scientometric methods are often criticized for lacking depth and context. As~\citet{igic_citation_2024} notes: 
\begin{quote}
``Ideally, evaluating scientific impact would involve reading each publication.'' 
\end{quote}

This sentiment highlights a fundamental challenge: while large-scale quantitative metrics offer breadth, they often miss the nuanced understanding that comes from close reading. Igić further advocates for combining classical qualitative assessment with scientometrics to achieve a more comprehensive evaluation.

\textit{Compare} addresses this challenge by enabling the automatic comparison of institutional research impact. Instead of relying solely on numerical indicators, it retrieves relevant literature and supports qualitative comparison of research contributions across institutions.
\section{\textit{Compare} Framework}\label{sec:compare}
\textit{Compare} is a modular framework designed to assist researchers, research analysts, and repository managers in conducting qualitative comparisons of scientific contributions at the publication or institutional level. Users interact with the system by submitting a natural language question, optionally accompanied by a scientific paper. Based on the prompt's intent, the system selects and executes one of several predefined analytical pipelines to generate a structured, citation-supported response. 

\textit{Compare} is built with Python 3.11, Flask 3.1, Streamlit, and LlamaIndex, integrating both publication full-text and metadata-centric data sources to support diverse analysis scenarios.

\subsection{Data Sources}
Currently, \textit{Compare} integrates two open-access research infrastructure APIs, from which it dynamically retrieves documents to populate its pipeline:

\begin{itemize}
    \item \textbf{CORE}~\cite{knoth_core_2012, knoth_core_2023} is a large-scale scholarly publications aggregation platform, offering access to the full texts of scientific publications, allowing for detailed content analysis.
    \item \textbf{OpenAlex}~\cite{priem_openalex_2022} is an open catalog of scholarly works, authors, institutions, and topics. It provides rich metadata (authors, affiliations, citations, etc.) for scholarly outputs, enabling high-level comparative analyses.
\end{itemize}

\subsection{Workflow}

\begin{figure}[tb]
    \centering
    \includegraphics[width=0.75\linewidth]{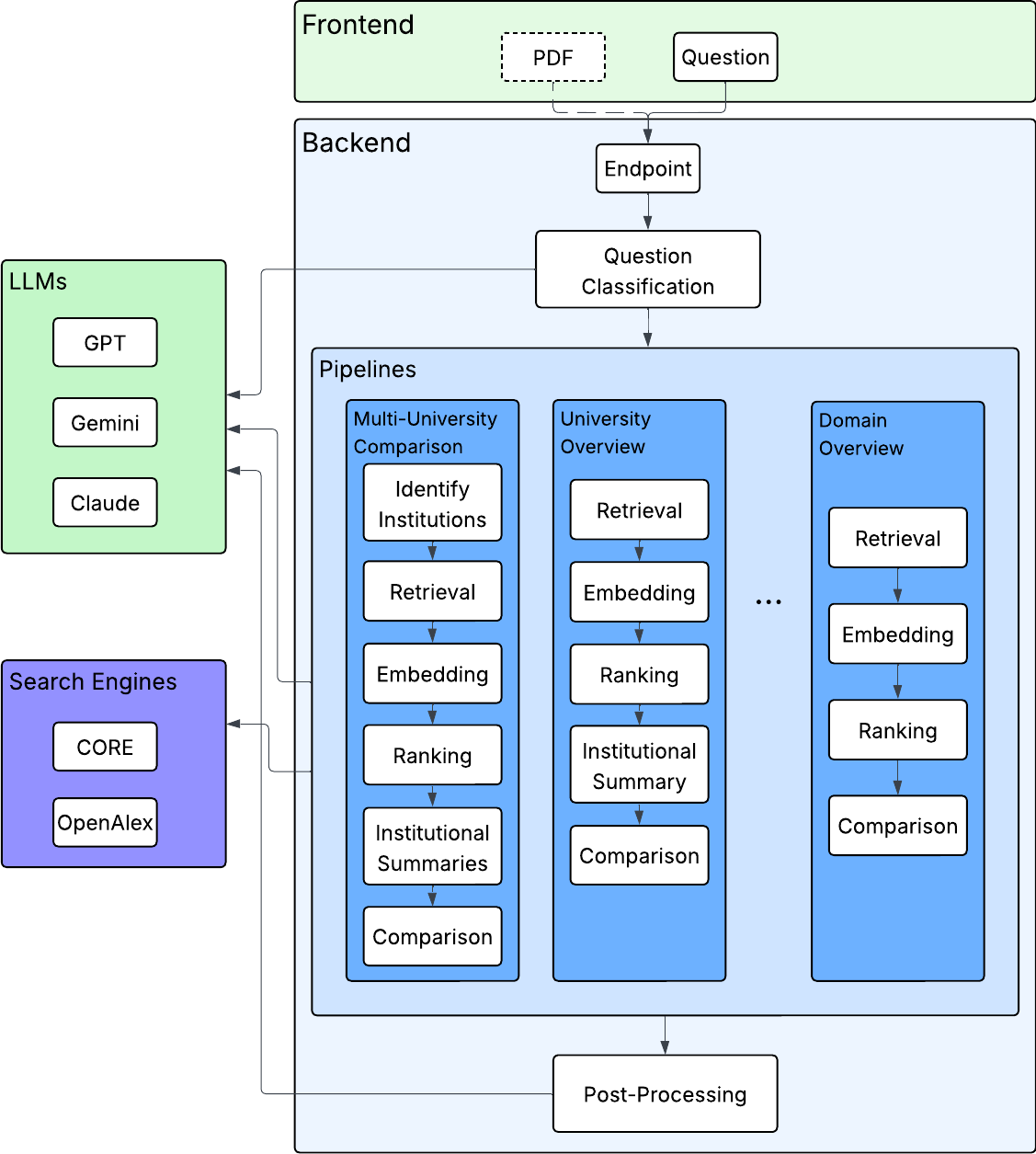}
    \caption{Overview of the \textit{Compare} framework, providing institutional and publication-level comparisons. Dotted elements are optional.}
    \label{fig:pipeline}
\end{figure}

Figure~\ref{fig:pipeline} illustrates the core workflow of the \textit{Compare} framework. Users begin by submitting a natural language question, optionally accompanied by a scientific publication in PDF format. The system preprocesses the input and classifies the query into one of several supported categories displayed in Table~\ref{tab:usecases}, namely: \textit{University Overview}, \textit{University Comparison}, \textit{Multi-University Comparison, }\textit{Domain Overview}, \textit{Paper Comparison}, or \textit{Paper QA}. 

Based on this classification, an appropriate pipeline is selected to fulfill the information need.

Each pipeline retrieves documents from the relevant data sources: OpenAlex or CORE. OpenAlex is used for metadata-level analysis, including author and institutional information, while CORE enables access to full-text content for deeper document-level comparisons. If available in the query text, metadata filters such as institution, country or publication year are applied to constrain the result set.

Once the relevant documents are retrieved, they are embedded using vector representations. Candidate documents are then ranked based on their relevance to the query. In institution-level pipelines, the system organizes the documents by affiliation and generates summaries for each institution. These summaries are compared to each other, taking into account the user’s original question.

In a final step, the system performs postprocessing to ensure output quality. This includes refining citation numbering and removing unused references from the output. The result is a structured, citation-grounded qualitative comparison tailored to the query.

\begin{figure}[tb]
    \centering
    \includegraphics[width=\linewidth]{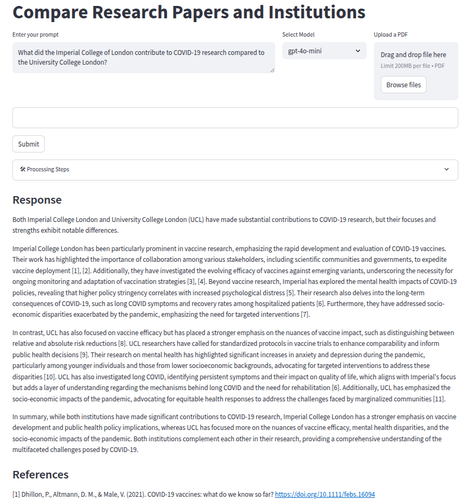}
    \caption{A screenshot of \textit{Compare} displaying qualitative analysis for a user query comparing the COVID-19 research of Imperial College London and University College London.}
    \label{fig:screenshot}
\end{figure}

\begin{figure}[tb]
    \centering
    \includegraphics[width=0.9\linewidth]{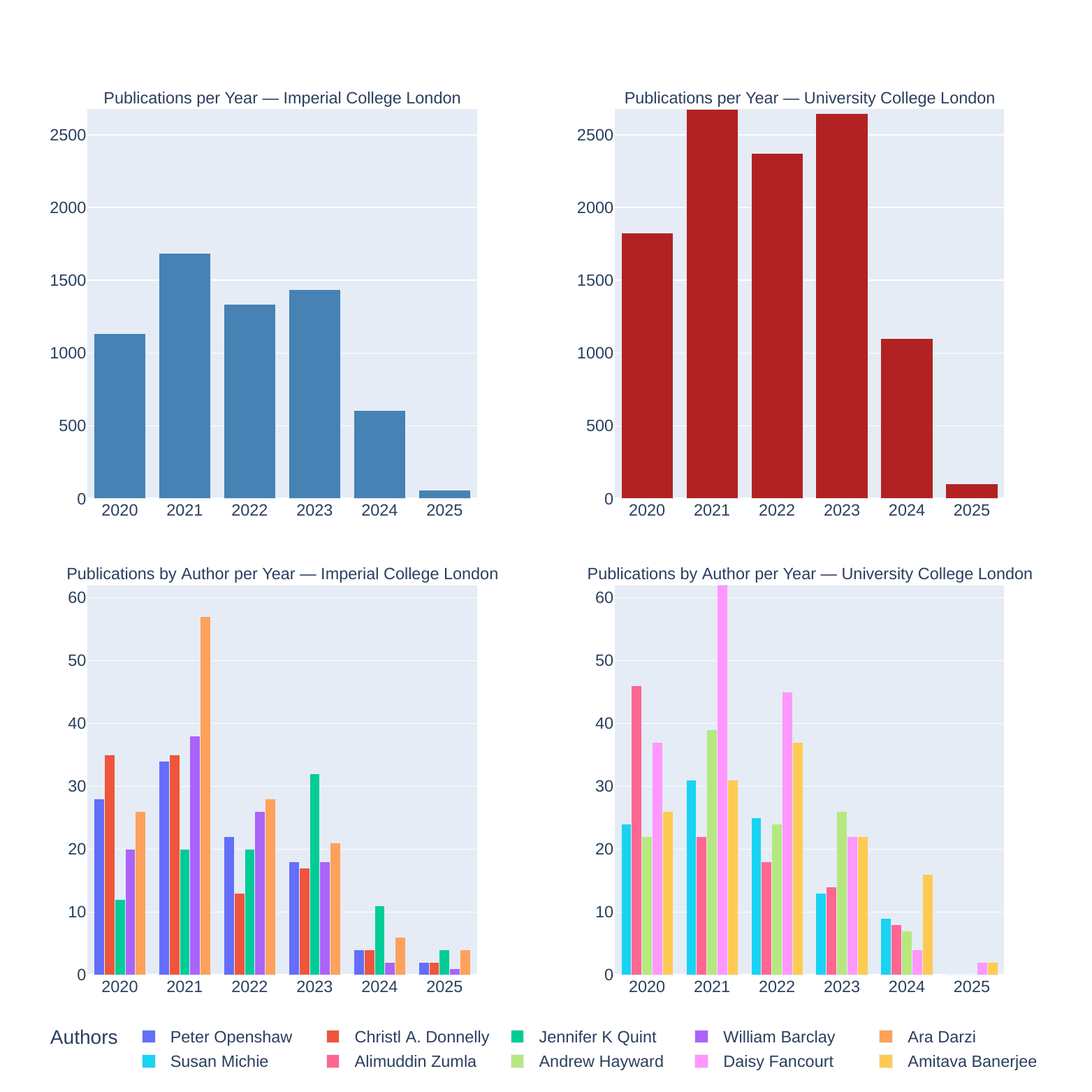}
    \caption{Quantitative analysis for the same COVID-19 research comparison shown in Figure~\ref{fig:screenshot}, including publication trends and key authors per institution.}
    \label{fig:diagrams}
\end{figure}

Figures~\ref{fig:screenshot} and~\ref{fig:diagrams} illustrate the outputs for the same query, which asks to compare the contributions in COVID-19 research between Imperial College London and University College London. While Figure~\ref{fig:screenshot} shows the qualitative comparison, Figure~\ref{fig:diagrams} complements it with quantitative visualizations.

Together with the generated text, all the used sources are listed with their DOIs and authors, as displayed in Figure~\ref{fig:screenshot}. 

In Figure~\ref{fig:diagrams} two of the sample output plots are displayed. The first row of plots shows the publications over time for the given universities, and the second pair of plots showcases the most relevant researchers for the given topic at the university with their publication numbers. 

In Table~\ref{tab:usecases} we describe our six supported use cases, with a short description of their pipelines, as well as an example query.

\begin{table*}[tb]
\small
\centering
\caption{Overview of currently supported use cases in the \textit{Compare} framework. Each use case maps to a specific query type and corresponding processing pipeline. All use cases except for `Paper QA' \textcolor{filtercolor}{support filters}.}
\label{tab:usecases}
\begin{tabular}{@{}p{2.6cm}p{2.8cm}p{5.2cm}p{4.7cm}@{}}
\toprule
\textbf{Use Case} & \textbf{Input} & \textbf{System Action} & \textbf{Example Query} \\
\midrule
\textbf{University Overview} & \textcolor{unicolor}{One University} + \textcolor{topiccolor}{topic} &
Retrieves publications affiliated with the university and summarizes key contributions. &
What has the \textcolor{unicolor}{University of Aberdeen} contributed to \textcolor{topiccolor}{Medical Natural Language Processing}? \\
\addlinespace[0.5ex]

\textbf{University Comparison} & \textcolor{biunicolor}{Two universities} + \textcolor{topiccolor}{topic} &
Retrieves and summarizes documents per institution, then generates a comparative synthesis. &
What has the \textcolor{biunicolor}{University of Aberdeen} contributed to \textcolor{topiccolor}{Information Retrieval} compared to the \textcolor{biunicolor}{University of Edinburgh}? \\
\addlinespace[0.5ex]

\textbf{Multi-University Comparison} & \textcolor{unicolor}{One university} + \textcolor{topiccolor}{topic} &
Identifies top 3 institutions by activity, compares them with the specified one. &
What has the \textcolor{unicolor}{University of Aberdeen} contributed to \textcolor{topiccolor}{Information Retrieval} compared to universities \textcolor{filtercolor}{in GB}? \\
\addlinespace[0.5ex]

\textbf{Domain Overview} & \textcolor{topiccolor}{Topic only} &
Retrieves full-text documents from CORE and summarizes the topic across the field. &
What are the advances in \textcolor{topiccolor}{multimodal learning} \textcolor{filtercolor}{since 2023}? \\
\addlinespace[0.5ex]

\textbf{Paper Comparison} & PDF + \textcolor{topiccolor}{topic}/\textcolor{aspectcolor}{aspect} &
Analyzes a paper’s aspect (e.g., datasets, methods) and compares with similar publications. &
How does the paper compare to other relevant literature \textcolor{aspectcolor}{based on its used models}? \\
\addlinespace[0.5ex]

\textbf{Paper QA} & PDF + \textcolor{questioncolor}{question} &
Answers the question using only the content of the provided paper. &
\textcolor{questioncolor}{What evaluation metrics are used in this paper?} \\
\bottomrule

\end{tabular}
\end{table*}

\section{Discussion}\label{sec:discussion}
\subsection{Data Quality}
Despite ongoing efforts to collect and harmonize metadata across a wide range of publication formats and venues, significant limitations in data quality remain. Many research contributions are still not openly accessible, and even when available, their metadata is often incomplete or inconsistent. Platforms like CORE and OpenAlex are working to maximize data availability, but gaps persist. For example, Priem~\cite{priem_openalex_2022} reports that only around 50\% of records in OpenAlex include DOIs. In our own analysis, we found that approximately 91 million publications in OpenAlex contain both an abstract and a DOI (34\% of the OpenAlex corpus), and 61 million (23\% of the OpenAlex corpus) of these also include institutional affiliation information\footnote{Retrieved via \url{https://api.openalex.org/works?filter=has_doi:true,has_abstract:true,authorships.affiliations.institution_ids:!null}}. \textit{Compare} therefore only uses 61 million of the 267 million records in the OpenAlex corpus.

Improving coverage by incorporating institutional repositories with higher-quality metadata could significantly enhance the performance and reliability of our retrieval pipelines.

\subsection{System Bias}
Our framework is subject to several types of bias. Selection bias is introduced when relevant works are missing due to them being excluded because of incomplete metadata or a lack of open access. Model bias is introduced to our framework as LLM occasionally hallucinate or exaggerate differences. Presentation bias occurs when generated text makes uncertain claims appear authoritative, especially if citations are imperfectly matched to the claims.

While these biases cannot be fully eliminated, we mitigate their impact by grounding outputs in retrieved sources and ensuring completeness of metadata.

\subsection{Expert Feedback}
To gain preliminary insights into the usefulness of our tool, we conducted informal evaluations with four researchers and research analysts from TU Wien, the Open University, Imperial College London, and the University of Birmingham. We demonstrated the core features of \textit{Compare}, particularly the generated summaries. We invited participants to reflect on the accuracy, relevance, and clarity of summaries.
After the demonstration, all participants got access to our framework, and were invited to explore the system themselves and send feedback via a 1–5 rating scale, with the option to leave textual comments. The feedback obtained so far is positive: all four participants identified potential ways in which \textit{Compare} could enhance their working routines. However, they also noted limitations - in particular, occasional incorrect citations (especially for \textit{gemini-2.0-flash}) and summaries that sometimes lack depth.
We consider this initial feedback a valuable checkpoint and a plan to conduct more structured user studies in future work to systematically assess the effectiveness and usability of the system.

\section{Conclusion}\label{sec:conclusion}
We presented \textit{Compare}, a RAG-based framework designed to streamline the exploration and analysis of scientific publications. Specifically, \textit{Compare} assists users in identifying research overlaps for scientific institutions and investigating the contributions of individual publications. Our framework provides a much-needed tool to navigate the increasingly complex landscape of academic literature.

A key strength of our framework is that it can be readily extended and adapted to diverse organizational needs, by integrating additional data sources and tailoring the framework to precise information requirements. 

Future work will focus on enhancing the framework's capabilities. We plan to facilitate a more straightforward integration of additional data sources and workflows, further amplifying the benefits for researchers, research analysts and repository managers.
Furthermore, we plan to refine our RAG pipeline to enable more advanced, agent and feedback-driven question answering, allowing for more nuanced and interactive explorations.
Finally, we intend to contribute LlamaIndex data loaders for CORE and OpenAlex to the community, promoting broader accessibility and utility of our work.
Our demo is available online\footnote{\url{https://compare.ds-ifs.tuwien.ac.at}}, as well as our source code\footnote{\url{https://github.com/MoritzStaudinger/compare}}.

\begin{acks}
We are grateful for the funding received from Open University through the REF2029 Student funding, which enabled Moritz to undertake this research in cooperation with the Open University. 
A special thanks also goes to Suchetha for her generosity in sharing her desk, making this work literally possible from a comfortable spot!
\end{acks}

\appendix
\section*{Use of Generative AI}
The authors used GPT-4 and Gemini 2.0-flash to paraphrase selected sentences and enhance the reading flow of the manuscript. All rewritten passages were carefully reviewed, edited, and approved by the authors. The core ideas, content, and interpretations remain entirely the original work of the authors.
\bibliographystyle{ACM-Reference-Format}
\bibliography{references}


\begin{thebibliography}{24}

%%% the Web address.

\ifx \showCODEN    \undefined \def \showCODEN     #1{\unskip}     \fi
\ifx \showISBNx    \undefined \def \showISBNx     #1{\unskip}     \fi
\ifx \showISBNxiii \undefined \def \showISBNxiii  #1{\unskip}     \fi
\ifx \showISSN     \undefined \def \showISSN      #1{\unskip}     \fi
\ifx \showLCCN     \undefined \def \showLCCN      #1{\unskip}     \fi
\ifx \shownote     \undefined \def \shownote      #1{#1}          \fi
\ifx \showarticletitle \undefined \def \showarticletitle #1{#1}   \fi
\ifx \showURL      \undefined \def \showURL       {\relax}        \fi
% invisible to TeX
\providecommand\bibfield[2]{#2}
\providecommand\bibinfo[2]{#2}
\providecommand\natexlab[1]{#1}
\providecommand\showeprint[2][]{arXiv:#2}

\bibitem[Agarwal et~al\mbox{.}(2025)]%
        {agarwal_litllms_2025}
\bibfield{author}{\bibinfo{person}{Shubham Agarwal}, \bibinfo{person}{Gaurav Sahu}, \bibinfo{person}{Abhay Puri}, \bibinfo{person}{Issam~H. Laradji}, \bibinfo{person}{Krishnamurthy~DJ Dvijotham}, \bibinfo{person}{Jason Stanley}, \bibinfo{person}{Laurent Charlin}, {and} \bibinfo{person}{Christopher Pal}.} \bibinfo{year}{2025}\natexlab{}.
\newblock \bibinfo{title}{{LitLLMs}, {LLMs} for {Literature} {Review}: {Are} we there yet?}
\newblock
\href{https://doi.org/10.48550/arXiv.2412.15249}{doi:\nolinkurl{10.48550/arXiv.2412.15249}}
\newblock
\shownote{arXiv:2412.15249 [cs]}.


\bibitem[Asai et~al\mbox{.}(2024)]%
        {asai_openscholar_2024}
\bibfield{author}{\bibinfo{person}{Akari Asai}, \bibinfo{person}{Jacqueline He}, \bibinfo{person}{Rulin Shao}, \bibinfo{person}{Weijia Shi}, \bibinfo{person}{Amanpreet Singh}, \bibinfo{person}{Joseph~Chee Chang}, \bibinfo{person}{Kyle Lo}, \bibinfo{person}{Luca Soldaini}, \bibinfo{person}{Sergey Feldman}, \bibinfo{person}{Mike D'arcy}, \bibinfo{person}{David Wadden}, \bibinfo{person}{Matt Latzke}, \bibinfo{person}{Minyang Tian}, \bibinfo{person}{Pan Ji}, \bibinfo{person}{Shengyan Liu}, \bibinfo{person}{Hao Tong}, \bibinfo{person}{Bohao Wu}, \bibinfo{person}{Yanyu Xiong}, \bibinfo{person}{Luke Zettlemoyer}, \bibinfo{person}{Graham Neubig}, \bibinfo{person}{Dan Weld}, \bibinfo{person}{Doug Downey}, \bibinfo{person}{Wen-tau Yih}, \bibinfo{person}{Pang~Wei Koh}, {and} \bibinfo{person}{Hannaneh Hajishirzi}.} \bibinfo{year}{2024}\natexlab{}.
\newblock \bibinfo{title}{{OpenScholar}: {Synthesizing} {Scientific} {Literature} with {Retrieval}-augmented {LMs}}.
\newblock
\href{https://doi.org/10.48550/arXiv.2411.14199}{doi:\nolinkurl{10.48550/arXiv.2411.14199}}
\newblock
\shownote{arXiv:2411.14199 [cs]}.


\bibitem[Auer et~al\mbox{.}(2023)]%
        {auer_sciqa_2023}
\bibfield{author}{\bibinfo{person}{Sören Auer}, \bibinfo{person}{Dante A.~C. Barone}, \bibinfo{person}{Cassiano Bartz}, \bibinfo{person}{Eduardo~G. Cortes}, \bibinfo{person}{Mohamad~Yaser Jaradeh}, \bibinfo{person}{Oliver Karras}, \bibinfo{person}{Manolis Koubarakis}, \bibinfo{person}{Dmitry Mouromtsev}, \bibinfo{person}{Dmitrii Pliukhin}, \bibinfo{person}{Daniil Radyush}, \bibinfo{person}{Ivan Shilin}, \bibinfo{person}{Markus Stocker}, {and} \bibinfo{person}{Eleni Tsalapati}.} \bibinfo{year}{2023}\natexlab{}.
\newblock \showarticletitle{The {SciQA} {Scientific} {Question} {Answering} {Benchmark} for {Scholarly} {Knowledge}}.
\newblock \bibinfo{journal}{\emph{Scientific Reports}} \bibinfo{volume}{13}, \bibinfo{number}{1} (\bibinfo{date}{May} \bibinfo{year}{2023}), \bibinfo{pages}{7240}.
\newblock
\showISSN{2045-2322}
\href{https://doi.org/10.1038/s41598-023-33607-z}{doi:\nolinkurl{10.1038/s41598-023-33607-z}}
\newblock
\shownote{Publisher: Nature Publishing Group}.


\bibitem[Bornmann et~al\mbox{.}(2021)]%
        {bornmann_growth_2021}
\bibfield{author}{\bibinfo{person}{Lutz Bornmann}, \bibinfo{person}{Robin Haunschild}, {and} \bibinfo{person}{Rüdiger Mutz}.} \bibinfo{year}{2021}\natexlab{}.
\newblock \showarticletitle{Growth rates of modern science: a latent piecewise growth curve approach to model publication numbers from established and new literature databases}.
\newblock \bibinfo{journal}{\emph{Humanities and Social Sciences Communications}} \bibinfo{volume}{8}, \bibinfo{number}{1} (\bibinfo{date}{Oct.} \bibinfo{year}{2021}), \bibinfo{pages}{1--15}.
\newblock
\showISSN{2662-9992}
\href{https://doi.org/10.1057/s41599-021-00903-w}{doi:\nolinkurl{10.1057/s41599-021-00903-w}}
\newblock
\shownote{Publisher: Palgrave}.


\bibitem[Igić(2024)]%
        {igic_citation_2024}
\bibfield{author}{\bibinfo{person}{Rajko Igić}.} \bibinfo{year}{2024}\natexlab{}.
\newblock \showarticletitle{Citation metrics and scientometrics}.
\newblock \bibinfo{journal}{\emph{Biomolecules and Biomedicine}} \bibinfo{volume}{24}, \bibinfo{number}{2} (\bibinfo{date}{April} \bibinfo{year}{2024}), \bibinfo{pages}{434--435}.
\newblock
\showISSN{2831-0896}
\href{https://doi.org/10.17305/bb.2023.10233}{doi:\nolinkurl{10.17305/bb.2023.10233}}


\bibitem[Kachooei and Ebrahimzadeh(2022)]%
        {kachooei_editorial_2022}
\bibfield{author}{\bibinfo{person}{Amir~R. Kachooei} {and} \bibinfo{person}{Mohammad~H. Ebrahimzadeh}.} \bibinfo{year}{2022}\natexlab{}.
\newblock \showarticletitle{Editorial: {What} {Is} {Peer} {Review}?}
\newblock \bibinfo{journal}{\emph{Archives of Bone and Joint Surgery}} \bibinfo{volume}{10}, \bibinfo{number}{1} (\bibinfo{date}{Jan.} \bibinfo{year}{2022}), \bibinfo{pages}{1--2}.
\newblock
\showISSN{2345-4644}
\href{https://doi.org/10.22038/abjs.2022.19585}{doi:\nolinkurl{10.22038/abjs.2022.19585}}


\bibitem[Knoth et~al\mbox{.}(2023)]%
        {knoth_core_2023}
\bibfield{author}{\bibinfo{person}{Petr Knoth}, \bibinfo{person}{Drahomira Herrmannova}, \bibinfo{person}{Matteo Cancellieri}, \bibinfo{person}{Lucas Anastasiou}, \bibinfo{person}{Nancy Pontika}, \bibinfo{person}{Samuel Pearce}, \bibinfo{person}{Bikash Gyawali}, {and} \bibinfo{person}{David Pride}.} \bibinfo{year}{2023}\natexlab{}.
\newblock \showarticletitle{{CORE}: {A} {Global} {Aggregation} {Service} for {Open} {Access} {Papers}}.
\newblock \bibinfo{journal}{\emph{Scientific Data}} \bibinfo{volume}{10}, \bibinfo{number}{1} (\bibinfo{date}{June} \bibinfo{year}{2023}), \bibinfo{pages}{366}.
\newblock
\showISSN{2052-4463}
\href{https://doi.org/10.1038/s41597-023-02208-w}{doi:\nolinkurl{10.1038/s41597-023-02208-w}}
\newblock
\shownote{Publisher: Nature Publishing Group}.


\bibitem[Knoth and Zdrahal(2012)]%
        {knoth_core_2012}
\bibfield{author}{\bibinfo{person}{Petr Knoth} {and} \bibinfo{person}{Zdenek Zdrahal}.} \bibinfo{year}{2012}\natexlab{}.
\newblock \showarticletitle{{CORE}: three access levels to underpin open access}.
\newblock \bibinfo{journal}{\emph{D-Lib Magazine}} \bibinfo{volume}{18}, \bibinfo{number}{11/12} (\bibinfo{year}{2012}).
\newblock
\showISSN{1082-9873}
\href{https://doi.org/10.1045/november2012-knoth}{doi:\nolinkurl{10.1045/november2012-knoth}}
\newblock
\shownote{Number: 11/12}.


\bibitem[Kulkarni et~al\mbox{.}(2020)]%
        {kulkarni_aquamuse_2020}
\bibfield{author}{\bibinfo{person}{Sayali Kulkarni}, \bibinfo{person}{Sheide Chammas}, \bibinfo{person}{Wan Zhu}, \bibinfo{person}{Fei Sha}, {and} \bibinfo{person}{Eugene Ie}.} \bibinfo{year}{2020}\natexlab{}.
\newblock \bibinfo{title}{{AQuaMuSe}: {Automatically} {Generating} {Datasets} for {Query}-{Based} {Multi}-{Document} {Summarization}}.
\newblock
\href{https://doi.org/10.48550/arXiv.2010.12694}{doi:\nolinkurl{10.48550/arXiv.2010.12694}}
\newblock
\shownote{arXiv:2010.12694 [cs]}.


\bibitem[Kusa et~al\mbox{.}(2023)]%
        {kusa_cruise-screening_2023}
\bibfield{author}{\bibinfo{person}{Wojciech Kusa}, \bibinfo{person}{Petr Knoth}, {and} \bibinfo{person}{Allan Hanbury}.} \bibinfo{year}{2023}\natexlab{}.
\newblock \showarticletitle{{CRUISE}-{Screening}: {Living} {Literature} {Reviews} {Toolbox}}. In \bibinfo{booktitle}{\emph{Proceedings of the 32nd {ACM} {International} {Conference} on {Information} and {Knowledge} {Management}}}. \bibinfo{publisher}{ACM}, \bibinfo{address}{Birmingham United Kingdom}, \bibinfo{pages}{5071--5075}.
\newblock
\showISBNx{9798400701245}
\href{https://doi.org/10.1145/3583780.3614736}{doi:\nolinkurl{10.1145/3583780.3614736}}


\bibitem[Lewis et~al\mbox{.}(2020)]%
        {lewis_retrieval-augmented_2020}
\bibfield{author}{\bibinfo{person}{Patrick Lewis}, \bibinfo{person}{Ethan Perez}, \bibinfo{person}{Aleksandra Piktus}, \bibinfo{person}{Fabio Petroni}, \bibinfo{person}{Vladimir Karpukhin}, \bibinfo{person}{Naman Goyal}, \bibinfo{person}{Heinrich Küttler}, \bibinfo{person}{Mike Lewis}, \bibinfo{person}{Wen-tau Yih}, \bibinfo{person}{Tim Rocktäschel}, \bibinfo{person}{Sebastian Riedel}, {and} \bibinfo{person}{Douwe Kiela}.} \bibinfo{year}{2020}\natexlab{}.
\newblock \showarticletitle{Retrieval-augmented generation for knowledge-intensive {NLP} tasks}. In \bibinfo{booktitle}{\emph{Proceedings of the 34th {International} {Conference} on {Neural} {Information} {Processing} {Systems}}} \emph{(\bibinfo{series}{{NIPS} '20})}. \bibinfo{publisher}{Curran Associates Inc.}, \bibinfo{address}{Red Hook, NY, USA}, \bibinfo{pages}{9459--9474}.
\newblock
\showISBNx{978-1-7138-2954-6}


\bibitem[Liu et~al\mbox{.}(2024)]%
        {liu_querysum_2024}
\bibfield{author}{\bibinfo{person}{Yushan Liu}, \bibinfo{person}{Zili Wang}, {and} \bibinfo{person}{Ruifeng Yuan}.} \bibinfo{year}{2024}\natexlab{}.
\newblock \showarticletitle{{QuerySum}: {A} {Multi}-{Document} {Query}-{Focused} {Summarization} {Dataset} {Augmented} with {Similar} {Query} {Clusters}}.
\newblock \bibinfo{journal}{\emph{Proceedings of the AAAI Conference on Artificial Intelligence}} \bibinfo{volume}{38}, \bibinfo{number}{17} (\bibinfo{date}{March} \bibinfo{year}{2024}), \bibinfo{pages}{18725--18732}.
\newblock
\showISSN{2374-3468}
\href{https://doi.org/10.1609/aaai.v38i17.29836}{doi:\nolinkurl{10.1609/aaai.v38i17.29836}}
\newblock
\shownote{Number: 17}.


\bibitem[Lu et~al\mbox{.}(2020)]%
        {lu_multi-xscience_2020}
\bibfield{author}{\bibinfo{person}{Yao Lu}, \bibinfo{person}{Yue Dong}, {and} \bibinfo{person}{Laurent Charlin}.} \bibinfo{year}{2020}\natexlab{}.
\newblock \showarticletitle{Multi-{XScience}: {A} {Large}-scale {Dataset} for {Extreme} {Multi}-document {Summarization} of {Scientific} {Articles}}. In \bibinfo{booktitle}{\emph{Proceedings of the 2020 {Conference} on {Empirical} {Methods} in {Natural} {Language} {Processing} ({EMNLP})}}, \bibfield{editor}{\bibinfo{person}{Bonnie Webber}, \bibinfo{person}{Trevor Cohn}, \bibinfo{person}{Yulan He}, {and} \bibinfo{person}{Yang Liu}} (Eds.). \bibinfo{publisher}{Association for Computational Linguistics}, \bibinfo{address}{Online}, \bibinfo{pages}{8068--8074}.
\newblock
\href{https://doi.org/10.18653/v1/2020.emnlp-main.648}{doi:\nolinkurl{10.18653/v1/2020.emnlp-main.648}}


\bibitem[Luo et~al\mbox{.}(2022)]%
        {luo_systematic_2022}
\bibfield{author}{\bibinfo{person}{Wenkai Luo}, \bibinfo{person}{Malindu Sandanayake}, \bibinfo{person}{Lei Hou}, \bibinfo{person}{Yongtao Tan}, {and} \bibinfo{person}{Guomin Zhang}.} \bibinfo{year}{2022}\natexlab{}.
\newblock \showarticletitle{A systematic review of green construction research using scientometrics methods}.
\newblock \bibinfo{journal}{\emph{Journal of Cleaner Production}}  \bibinfo{volume}{366} (\bibinfo{date}{Sept.} \bibinfo{year}{2022}), \bibinfo{pages}{132710}.
\newblock
\showISSN{0959-6526}
\href{https://doi.org/10.1016/j.jclepro.2022.132710}{doi:\nolinkurl{10.1016/j.jclepro.2022.132710}}


\bibitem[Nema et~al\mbox{.}(2017)]%
        {nema_diversity_2017}
\bibfield{author}{\bibinfo{person}{Preksha Nema}, \bibinfo{person}{Mitesh~M. Khapra}, \bibinfo{person}{Anirban Laha}, {and} \bibinfo{person}{Balaraman Ravindran}.} \bibinfo{year}{2017}\natexlab{}.
\newblock \showarticletitle{Diversity driven attention model for query-based abstractive summarization}. In \bibinfo{booktitle}{\emph{Proceedings of the 55th {Annual} {Meeting} of the {Association} for {Computational} {Linguistics} ({Volume} 1: {Long} {Papers})}}, \bibfield{editor}{\bibinfo{person}{Regina Barzilay} {and} \bibinfo{person}{Min-Yen Kan}} (Eds.). \bibinfo{publisher}{Association for Computational Linguistics}, \bibinfo{address}{Vancouver, Canada}, \bibinfo{pages}{1063--1072}.
\newblock
\href{https://doi.org/10.18653/v1/P17-1098}{doi:\nolinkurl{10.18653/v1/P17-1098}}


\bibitem[Oyewola and Dada(2022)]%
        {oyewola_exploring_2022}
\bibfield{author}{\bibinfo{person}{David~Opeoluwa Oyewola} {and} \bibinfo{person}{Emmanuel~Gbenga Dada}.} \bibinfo{year}{2022}\natexlab{}.
\newblock \showarticletitle{Exploring machine learning: a scientometrics approach using bibliometrix and {VOSviewer}}.
\newblock \bibinfo{journal}{\emph{SN Applied Sciences}} \bibinfo{volume}{4}, \bibinfo{number}{5} (\bibinfo{date}{April} \bibinfo{year}{2022}), \bibinfo{pages}{143}.
\newblock
\showISSN{2523-3971}
\href{https://doi.org/10.1007/s42452-022-05027-7}{doi:\nolinkurl{10.1007/s42452-022-05027-7}}


\bibitem[Pilault et~al\mbox{.}(2020)]%
        {pilault_extractive_2020}
\bibfield{author}{\bibinfo{person}{Jonathan Pilault}, \bibinfo{person}{Raymond Li}, \bibinfo{person}{Sandeep Subramanian}, {and} \bibinfo{person}{Chris Pal}.} \bibinfo{year}{2020}\natexlab{}.
\newblock \showarticletitle{On {Extractive} and {Abstractive} {Neural} {Document} {Summarization} with {Transformer} {Language} {Models}}. In \bibinfo{booktitle}{\emph{Proceedings of the 2020 {Conference} on {Empirical} {Methods} in {Natural} {Language} {Processing} ({EMNLP})}}, \bibfield{editor}{\bibinfo{person}{Bonnie Webber}, \bibinfo{person}{Trevor Cohn}, \bibinfo{person}{Yulan He}, {and} \bibinfo{person}{Yang Liu}} (Eds.). \bibinfo{publisher}{Association for Computational Linguistics}, \bibinfo{address}{Online}, \bibinfo{pages}{9308--9319}.
\newblock
\href{https://doi.org/10.18653/v1/2020.emnlp-main.748}{doi:\nolinkurl{10.18653/v1/2020.emnlp-main.748}}


\bibitem[Pride et~al\mbox{.}(2023)]%
        {pride_core-gpt_2023}
\bibfield{author}{\bibinfo{person}{David Pride}, \bibinfo{person}{Matteo Cancellieri}, {and} \bibinfo{person}{Petr Knoth}.} \bibinfo{year}{2023}\natexlab{}.
\newblock \showarticletitle{{CORE}-{GPT}: {Combining} {Open} {Access} {Research} and {Large} {Language} {Models} for {Credible}, {Trustworthy} {Question} {Answering}}. In \bibinfo{booktitle}{\emph{Linking {Theory} and {Practice} of {Digital} {Libraries}}}, \bibfield{editor}{\bibinfo{person}{Omar Alonso}, \bibinfo{person}{Helena Cousijn}, \bibinfo{person}{Gianmaria Silvello}, \bibinfo{person}{Mónica Marrero}, \bibinfo{person}{Carla Teixeira~Lopes}, {and} \bibinfo{person}{Stefano Marchesin}} (Eds.). \bibinfo{publisher}{Springer Nature Switzerland}, \bibinfo{address}{Cham}, \bibinfo{pages}{146--159}.
\newblock
\showISBNx{978-3-031-43849-3}
\href{https://doi.org/10.1007/978-3-031-43849-3_13}{doi:\nolinkurl{10.1007/978-3-031-43849-3_13}}


\bibitem[Priem et~al\mbox{.}(2022)]%
        {priem_openalex_2022}
\bibfield{author}{\bibinfo{person}{Jason Priem}, \bibinfo{person}{Heather Piwowar}, {and} \bibinfo{person}{Richard Orr}.} \bibinfo{year}{2022}\natexlab{}.
\newblock \bibinfo{title}{{OpenAlex}: {A} fully-open index of scholarly works, authors, venues, institutions, and concepts}.
\newblock
\urldef\tempurl%
\url{https://arxiv.org/abs/2205.01833v2}
\showURL{%
\tempurl}


\bibitem[Roy and Kundu(2024)]%
        {roy_review_2024}
\bibfield{author}{\bibinfo{person}{Prasenjeet Roy} {and} \bibinfo{person}{Suman Kundu}.} \bibinfo{year}{2024}\natexlab{}.
\newblock \showarticletitle{Review on {Query}-focused {Multi}-document {Summarization} ({QMDS}) with {Comparative} {Analysis}}.
\newblock \bibinfo{journal}{\emph{Comput. Surveys}} \bibinfo{volume}{56}, \bibinfo{number}{1} (\bibinfo{date}{Jan.} \bibinfo{year}{2024}), \bibinfo{pages}{1--38}.
\newblock
\showISSN{0360-0300, 1557-7341}
\href{https://doi.org/10.1145/3597299}{doi:\nolinkurl{10.1145/3597299}}


\bibitem[Singh et~al\mbox{.}(2025)]%
        {singh_ai2_2025}
\bibfield{author}{\bibinfo{person}{Amanpreet Singh}, \bibinfo{person}{Joseph~Chee Chang}, \bibinfo{person}{Chloe Anastasiades}, \bibinfo{person}{Dany Haddad}, \bibinfo{person}{Aakanksha Naik}, \bibinfo{person}{Amber Tanaka}, \bibinfo{person}{Angele Zamarron}, \bibinfo{person}{Cecile Nguyen}, \bibinfo{person}{Jena~D. Hwang}, \bibinfo{person}{Jason Dunkleberger}, \bibinfo{person}{Matt Latzke}, \bibinfo{person}{Smita Rao}, \bibinfo{person}{Jaron Lochner}, \bibinfo{person}{Rob Evans}, \bibinfo{person}{Rodney Kinney}, \bibinfo{person}{Daniel~S. Weld}, \bibinfo{person}{Doug Downey}, {and} \bibinfo{person}{Sergey Feldman}.} \bibinfo{year}{2025}\natexlab{}.
\newblock \bibinfo{title}{Ai2 {Scholar} {QA}: {Organized} {Literature} {Synthesis} with {Attribution}}.
\newblock
\urldef\tempurl%
\url{https://arxiv.org/abs/2504.10861v1}
\showURL{%
\tempurl}


\bibitem[Tang and Yang(2024)]%
        {tang_multihop-rag_2024}
\bibfield{author}{\bibinfo{person}{Yixuan Tang} {and} \bibinfo{person}{Yi Yang}.} \bibinfo{year}{2024}\natexlab{}.
\newblock \bibinfo{title}{{MultiHop}-{RAG}: {Benchmarking} {Retrieval}-{Augmented} {Generation} for {Multi}-{Hop} {Queries}}.
\newblock
\href{https://doi.org/10.48550/arXiv.2401.15391}{doi:\nolinkurl{10.48550/arXiv.2401.15391}}
\newblock
\shownote{arXiv:2401.15391 [cs]}.


\bibitem[Wang et~al\mbox{.}(2024)]%
        {wang_leave_2024}
\bibfield{author}{\bibinfo{person}{Minzheng Wang}, \bibinfo{person}{Longze Chen}, \bibinfo{person}{Fu Cheng}, \bibinfo{person}{Shengyi Liao}, \bibinfo{person}{Xinghua Zhang}, \bibinfo{person}{Bingli Wu}, \bibinfo{person}{Haiyang Yu}, \bibinfo{person}{Nan Xu}, \bibinfo{person}{Lei Zhang}, \bibinfo{person}{Run Luo}, \bibinfo{person}{Yunshui Li}, \bibinfo{person}{Min Yang}, \bibinfo{person}{Fei Huang}, {and} \bibinfo{person}{Yongbin Li}.} \bibinfo{year}{2024}\natexlab{}.
\newblock \showarticletitle{Leave {No} {Document} {Behind}: {Benchmarking} {Long}-{Context} {LLMs} with {Extended} {Multi}-{Doc} {QA}}. In \bibinfo{booktitle}{\emph{Proceedings of the 2024 {Conference} on {Empirical} {Methods} in {Natural} {Language} {Processing}}}, \bibfield{editor}{\bibinfo{person}{Yaser Al-Onaizan}, \bibinfo{person}{Mohit Bansal}, {and} \bibinfo{person}{Yun-Nung Chen}} (Eds.). \bibinfo{publisher}{Association for Computational Linguistics}, \bibinfo{address}{Miami, Florida, USA}, \bibinfo{pages}{5627--5646}.
\newblock
\href{https://doi.org/10.18653/v1/2024.emnlp-main.322}{doi:\nolinkurl{10.18653/v1/2024.emnlp-main.322}}


\bibitem[Yang et~al\mbox{.}(2018)]%
        {yang_hotpotqa_2018}
\bibfield{author}{\bibinfo{person}{Zhilin Yang}, \bibinfo{person}{Peng Qi}, \bibinfo{person}{Saizheng Zhang}, \bibinfo{person}{Yoshua Bengio}, \bibinfo{person}{William~W. Cohen}, \bibinfo{person}{Ruslan Salakhutdinov}, {and} \bibinfo{person}{Christopher~D. Manning}.} \bibinfo{year}{2018}\natexlab{}.
\newblock \bibinfo{title}{{HotpotQA}: {A} {Dataset} for {Diverse}, {Explainable} {Multi}-hop {Question} {Answering}}.
\newblock
\href{https://doi.org/10.48550/arXiv.1809.09600}{doi:\nolinkurl{10.48550/arXiv.1809.09600}}
\newblock
\shownote{arXiv:1809.09600 [cs]}.


\end{thebibliography}

\end{document}